# Thresholdless Nanoscale Coaxial Lasers

M. Khajavikhan[1*], A. Simic[1**], M. Katz[1**], J. H. Lee[1], B. Slutsky[1], A. Mizrahi[1], V. Lomakin[1] and, Y. Fainman[1]

[1]*Dept. of Electrical and Computer Engineering, University of California San Diego, 9500 Gilman Dr. La Jolla, California 92093-0407, USA*

[*] *Corresponding author, email: mercedeh@ucsd.edu,*

[**] *These authors contributed equally to this work*

**The generation of coherent radiation in nanostructures has attracted considerable interest in recent years owing both to the quantum electrodynamical effects that emerge in small volumes, and to their potential for future applications. The progress towards exploring this size regime, however, has been hindered by the lack of a systematic approach to scaling down the mode as the structures shrink in size and by the high threshold power required for lasing in small cavities. Here we present a new family of metamaterial-like nano-cavities consisting of metallic coaxial nanostructures that solves both of these problems. The proposed nanoscale coaxial cavities can operate as thresholdless lasers, and their size is not restricted by the mode cut-off. Using these new nano-cavities we have succeeded to demonstrate for the first time thresholdless lasing in a broadband gain medium, and the smallest room temperature, continuous wave, telecom laser to date. The nanoscale coaxial cavities are the first systematic step towards unveiling the potentials of ultra small quantum electrodynamical objects in which atom-field interactions generate new functionalities.**

The cavity quantum electro dynamical effects caused by the interaction of the matter and field at sub-wavelength structures have been the subject of intense research in recent years. One relatively well-studied area in which these effects are of primary importance is in the design of nanolasers. Nanolasers are an emerging field of science, with a wide range of applications from on-chip optical communication, ultra high resolution imaging, high throughput sensing, and single molecule spectroscopy, to the study of atom-field interaction in ultra small cavities. Research is directed at designing the "ultimate nanolaser": a scalable, low threshold, efficient source of radiation that operates at room-temperature, and occupies a small area on chip[5].

There are currently two main approaches in designing nanolasers. The first approach utilizes dielectric based structures. Dielectrics have low loss at optical frequencies and can be designed as multilayer stacks to provide strong optical feedback. Examples of lasers using such structures are vertical cavity surface emitting lasers (VCSEL)[6], micro/nano disks[7,8], and photonic bandgap lasers[8,9,10,11]. There are, however, several drawbacks in using dielectric-based nanolasers. They are either large in size or their mode extends far out of the gain region, and thus they exhibit poor gain-mode overlap. Furthermore, light confinement with dielectrics, either by total internal reflection or stopband reflection, entails inherent scalability limitation. The second approach in designing nanolasers uses metal in the cavity. In recent years, nanoscale metallic[12], plasmonic[13,14,15,16,17], and metallo-dielectric[18,19,20] cavities have shown to confine light in ultra-small volumes and to improve the gain-mode spatial overlap. Moreover, metal cavities offer better thermal management in comparison to dielectric cavities and are more suitable for electrical pumping. However, existing metal-based nanolasers require high threshold pump power because of the significant absorption loss of the metals at optical frequencies. In summary, the miniaturization of laser cavities either by dielectric or metallic structures faces two challenges: one mode scalability meaning that as the cavity size decreases, it may no longer support a lasing mode. Another challenge is the fact that as the size of the cavity shrinks, optical gain decreases faster than cavity losses, causing the lasing threshold to become very large and/or unattainable.

In this work, we propose a new approach to nano-cavity design that solves both issues of high threshold power and un-scalable modes. Firstly, by designing a cavity that supports the cut-off-less transverse electromagnetic (TEM) mode, sub-wavelength size nano-cavities with modes far smaller than the operating wavelength can be realized. Secondly, the high threshold of lasers can be reduced through proper utilization of Purcell effect[21] by designing a cavity that enhances the interaction of emitters of the active medium with the cavity electromagnetic modes, and therefore increases the coupling of spontaneous emission into the lasing mode. Ultimately, the threshold constraint can be completely eliminated by reaching so-called *thresholdless* lasing, which occurs when every photon emitted by the gain medium is funneled into the lasing mode[22].

Here we present a validation of the above approach by reporting the first demonstration of lasing in metal based nanoscale coaxial cavities. Inspired by the coaxial resonators used in the microwave domain[23], we present the smallest *telecom* nanolaser to date that operates in the *continuous wave* regime at *room temperature*. Furthermore, we designed a nanoscale coaxial cavity that fulfills the conditions for *thresholdless* lasing which we observe at cryogenic temperatures. This first demonstration of a truly thresholdless lasing from a broadband gain medium emphasizes the unique properties of nanoscale coaxial structures for the harnessing of electrodynamical effects at sub-wavelength structures. Furthermore, these structures can be envisioned as the building blocks of gain assisted quantum meta-materials[24,25].

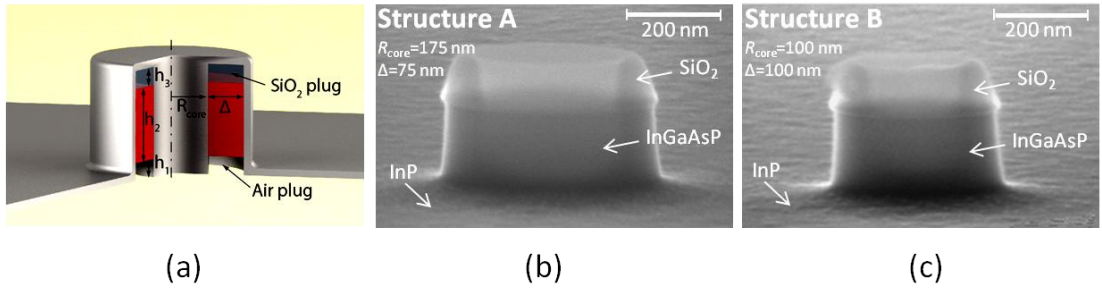

**Figure 1 | Nanoscale coaxial laser cavity.** (a) Schematic of a coaxial laser cavity. (b) and (c) SEM images of the constituent rings in Structure A and Structure B, respectively. The side view of the rings comprising the coaxial structures are seen. The rings consist of $SiO_2$ on top and quantum wells gain region underneath.

The coaxial laser cavity is portrayed in Figure 1(a). At the heart of the cavity lies a coaxial waveguide that is composed of a metallic rod enclosed by a metal coated semiconductor ring[26,27]. To form a cavity, the coaxial waveguide is capped at both ends with thin low index dielectric plugs. The upper plug is made of silicon dioxide and is covered by silver. The lower plug, which channels light out of the laser cavity and allows the pump beam to enter, is filled with air. The metal in the coaxial cavity is placed in direct contact with the semiconductor to ensure a large overlap between the mode and the emitters of the gain medium. In addition, the metallic coating serves as a heat sink.

To reduce the lasing threshold, the coaxial structures are carefully designed to maximize the benefits from the modification of the spontaneous emission due to the Purcell effect. Because of their small sizes, the modal content of the nanoscale coaxial cavities has intrinsic sparsity which is a key requirement to obtain high spontaneous emission coupling into the lasing mode. Their modal content can be further modified by tailoring the geometry, i.e. the radius of the core, the width of the ring, and the height of the low index plugs, bearing in mind that the number of modes that may participate in the lasing process is ultimately limited to those that are within the gain bandwidth of the semiconductor active material. The semiconductor gain medium used in this work is composed of six quantum wells of $In_{x=0.734}Ga_{1-x}As_{y=0.57}P_{1-y}$(10 nm thick)/ $In_{x=0.56}Ga_{1-x}As_{y=0.938}P_{1-y}$ (20 nm thick), resulting in a gain bandwidth that spans from 1.26 μm to 1.59 μm at room-temperature and from 1.27 μm to 1.53 μm at 4.5 K [28].

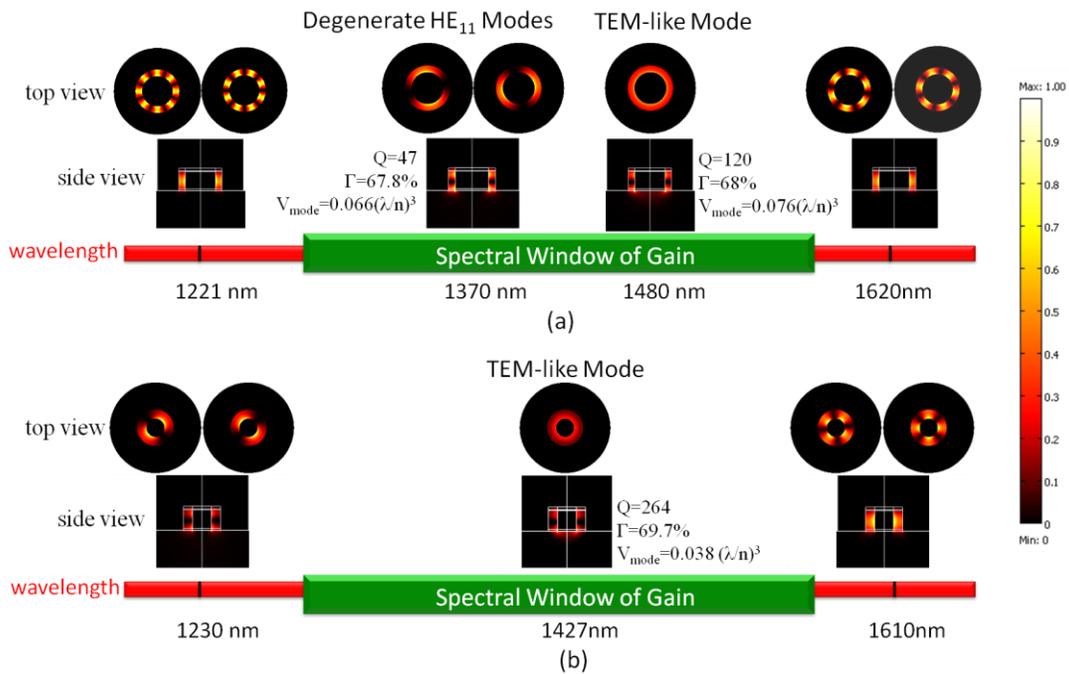

**Figure 2 | Electromagnetic simulation of nanoscale coaxial cavities.** (a) The modal spectrum of the cavity of Structure A at a temperature of 4.5 K. This cavity supports a pair of $HE_{11}$ degenerate modes and the fundamental TEM-like mode in the gain bandwidth. (b) The modal spectrum of the cavity of Structure B. This cavity only supports the fundamental TEM-like mode in the gain bandwidth of the quantum wells. In the figures, Q is the quality factor of the mode, Γ is the energy confinement factor[29], and $V_{mode}$ is the effective modal volume[29].

We consider below two different geometries based on the schematic of Figure 1(a). The first, which we refer to as *Structure A*, has an inner core radius of $R_{core}$=175 nm, gain medium ring of $\Delta$=75 nm, lower plug height of $h_1$= 20 nm, gain medium height of $h_2$=210 nm, and upper plug height of $h_3$=30 nm. *Structure B* is smaller in diameter having $R_{core}$=100 nm, and $\Delta$=100 nm. The heights of the plugs and gain medium are identical to those of Structure A. The two structures are fabricated using standard nanofabrication techniques. Figure 1 (b) and (c) show the SEM images of the constituent rings in Structure A and Structure B, respectively. For the details of the fabrication procedure see the supplementary information part 1.

In Figure 2 we present the modal content of the two structures at a temperature of 4.5 K, modeled using the Finite Element Method (FEM). Figure 2(a) shows that for Structure A the fundamental TEM-like mode and the two degenerate $HE_{11}$ modes are within the gain bandwidth of the active material. This simulation was also repeated for Structure A with room-temperature parameters (not shown). The two degenerate $HE_{11}$ modes are red shifted to 1400 nm, and exhibit a reduced quality factor of $Q \approx 35$ compared to $Q \approx 47$ at 4.5 K. The TEM-like mode is red shifted to 1520 nm with $Q \approx 53$ compared to $Q \approx 120$ at 4.5 K . For the details of the material constants at 4.5 K and at room-temperature see supplementary information part 2.

Structure B, as shown in Figure 2(b), supports only the fundamental TEM-like mode at a temperature of 4.5 K. The quality factor of this mode at $Q \approx 265$, is higher than that of Structure A. In general, the metal coating and the small aperture of the nanoscale coaxial cavity inhibit the gain emitters from coupling into the continuum of the free-space radiation modes [30,31]. Hence, the single mode cavity of Structure B exhibits very high spontaneous emission coupling factor (β→1), thereby operating very close to the ideal thresholdless laser [32]. This is in contrast to Structure A in which the spontaneous emission is distributed between three modes of the cavity.

It should be noted that the central wavelength of the modes is very sensitive to the size of the cavity. Variations as small as a few nanometers in diameter may cause a shift of tens of nanometers in the spectral location of the modes. This means that our existing fabrication and characterization tools do not allow us to predict the exact location of the modes in the output spectrum of the fabricated structures before performing the optical measurements.

The characterization of the nanoscale coaxial lasers was performed under optical pumping with a 1064 nm laser beam. We used continuous wave pumping up to an incident power of about 25 μW on the aperture of the cavities. For details of the measurement system see the supplementary information part 3.

Figure 3 shows the emission characteristics of the nanoscale coaxial laser of Structure A operating at 4.5 K (light-light curve in frame (a), and spectral evolution in frame (b)) and at room-temperature (light-light curve in frame (c) and spectral evolution in frame (d)). The light-light curves of Figure 3(a) and (c) show standard lasing behavior where spontaneous emission dominates at lower pump powers (referred to as the PL region) and stimulated emission is dominant at higher pump powers (referred to as the Lasing region). The PL and Lasing regions are connected through a pronounced transient region called amplified spontaneous emission (ASE) [33]. In Figures 3(b) and (d) the multiple peaks that are discerned at low pump powers reflect the modification of spontaneous emission spectrum by the cavity resonances depicted in Figure 2(a). The linewidth of the lasers shown in the insets of Figures 3(a) and (c) narrows with the inverse of the output power at lower pump levels (the solid trend line). This is in agreement with the well-known Schawlow-Townes formula for the laser operating below threshold [34]. Around threshold, the rapid increase of the gain-index coupling in semiconductor lasers slows down the narrowing of the linewidth, until carrier pinning resumes the Schawlow-Townes inverse power narrowing rate [35,36]. The mechanisms affecting the linewidth above threshold, especially for lasers with high spontaneous emission coupling to the lasing mode, are still a subject of research and addressing them is beyond the scope of this paper[37,38].

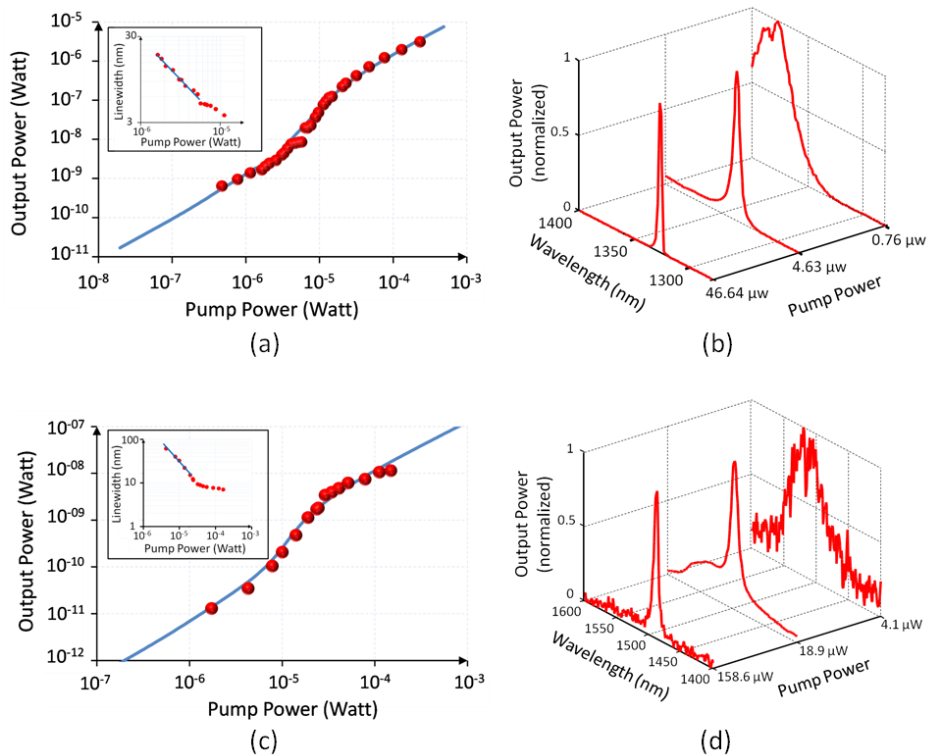

**Figure 3 | Optical characterization of nanoscale coaxial cavities, light-light curve, linewidth vs. pump power, and spectral evolution diagram for lasers with threshold.** Lasing in Structure A. (a) Light-light curve, (b) Spectral evolution at a temperature of 4.5 K, (c) Light-light curve, and (d) Spectral evolution at room-temperature. The pump power is calculated as the fraction of the power incident on the laser aperture. The solid curves in (a) and (c) are the best fit to the rate equation model. The insets show the linewidth vs. pump power. The resolution of the spectrometer was set to 3.3 nm.

A rate-equation model is adopted to study the dynamics of the photon-carriers interaction. For details of the rate-equation model see the supplementary information part 4. The light-light curves obtained from the rate equation model for the laser of Structure A are shown by solid blue lines in Figures 3(a) and (c). For the laser operating at 4.5 K, by fitting the experimental data to the rate equation model, we found that almost 20 percent of the spontaneous emission is coupled to the lasing mode (the mode with the highest quality factor). At room temperature, the surface and Auger non radiative recombination processes dominate, and as the carriers are lost through the non radiative channels, the ASE kink is more pronounced, and the threshold shifts to slightly higher pump powers.

Next, we measure the emission characteristics of Structure B. According to the electromagnetic analysis this structure operates as a thresholdless laser, as only one non-degenerate mode resides within the gain medium emission bandwidth. The emission characteristics of structure B at a temperature of 4.5 K are shown in Figure 4. The light-light curve of Figure 4(a) that follows a straight line with no pronounced kink agrees with the thresholdless lasing hypothesis[22]. The thresholdless behavior further manifests itself in the spectral evolution, seen in Figure 4(b), where a single narrow Lorentzian emission is obtained over the entire five-orders-of-magnitude range of pump power. This range spans from the first signal detected above the measurement system noise floor at 720 pW pump power, to the highest pump power of more than 100 µW. Since the emission spectrum of the gain medium is broad, the observed spectrum is attributed to the cavity mode, with the linewidth at low pump powers of about 5 nm that agrees well with the calculated quality factor (Q) of the TEM-like mode. The linewidth depicted in the inset of Figure 4(a) is almost constant and does not narrow inversely with the output power as predicted by Schawlow-Townes formula at lower pump levels, meaning that the linewidth shows no subthreshold behavior [37,38]. To the best of our knowledge, this linewidth behavior has never been reported in any other laser, and it may be the signature of thresholdless lasers. Finally, our rate equation model best fits the experimental data if 95 percent of the spontaneous emission is coupled to the lasing mode ($\beta=0.95$). At room temperature Structure B supports more than one mode and no longer fulfills the conditions for thresholdless lasing. In addition, excess non radiative recombination processes at room temperature makes it more difficult to distinguish and characterize thresholdless lasing.

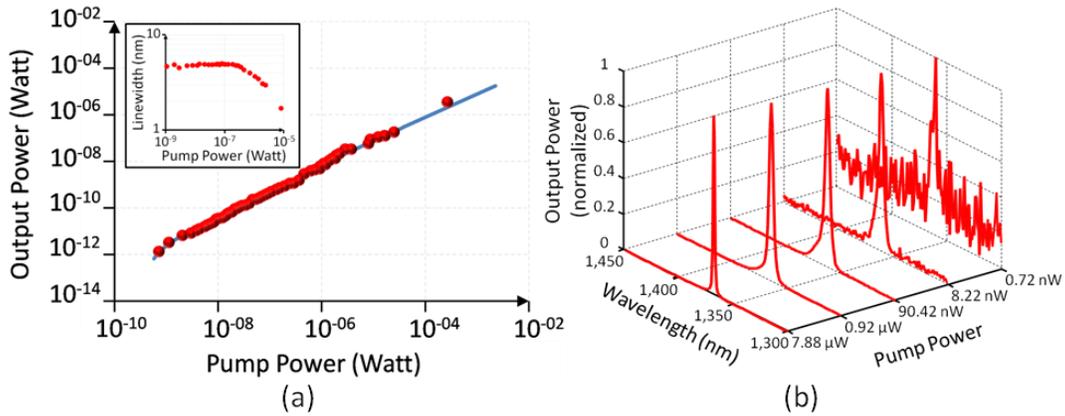

**Figure 4| Optical characterization of nanoscale coaxial cavities, light-light curve, linewidth vs. pump power, and spectral evolution diagram for thresholdless lasers.** Thresholdless lasing in Structure B. (a) Light-light curve at a temperature of 4.5 K, (b) Spectral evolution at a temperature of 4.5 K. The pump power is calculated as the fraction of the power incident on the laser aperture. The solid curve in (a) is the best fit to the rate equation model. The insets show the linewidth vs. pump power. The resolution of the spectrometer was set to 1.6 nm.

The thresholdless lasing in nanoscale coaxial cavities is in sharp contrast to the state-of-the-art high quality factor photonic bandgap structures. In the latter, near-thresholdless lasing is achieved in a gain medium which consists of quantum dots with spectrally narrowband emission, and relies extensively on tuning of the cavity mode to the center of the quantum dot emission spectrum[10]. In the former, thresholdless lasing in a broadband gain medium is achieved with a low quality factor single mode metal cavity.

In summary, with our nanoscale coaxial structures we have succeeded to demonstrate both room-temperature, continuous wave lasing and low temperature thresholdless lasing in a genuinely broadband semiconductor gain setting. Owing to the fundamental TEM-like mode with no cut-off, these cavities show ultra-small mode confinement, offer large mode-emitter overlap that results in an optimum utilization of the pump power, and provide multifold scalability. Further developments towards electrical pumping of thresholdless nanoscale coaxial lasers that can operate at room temperature are in progress.

The implication of our work is threefold. Firstly, the demonstrated nanoscale coaxial lasers have a great potential for future nano-photonic circuits on a chip. Secondly, thresholdless operation in tandem with scalability provide the first systematic approach toward the realization of quantum electro dynamical objects and functionalities, specifically, the realization of quantum meta-materials. Finally, this new family of resonators pave the way towards in-depth study of the unexplored physics of atom-field interaction, photon statistics, and carrier dynamics in ultra-small metallic structures.

**Acknowledgement**

The authors would like to thank the support from the Defense Advanced Research Projects Agency (DARPA), the National Science Foundation (NSF), the NSF Center for Integrated Access Networks (CIAN), the Cymer Corporation, and the U.S. Army Research Office. The authors would also like to thank the personnel of the UCSD Nano3 facilities for their help and support. M.Kh. would like to thank Professor Tara Javidi and Professor Jim Leger for helpful discussions.

**Author contributions**

The idea of thresholdless laser using nanoscale coaxial structures is conceived by M.Kh. The electromagnetic design, simulation, and analysis of the structures carried out by M.Kh., A.M and V.L. Fabrication of the devices was carried out by M.Kh. and J.H.L. The optical measurement were performed by A.S. and M.Kh. Rate equation model is developed by M.Ka. The optical characterization and analysis of laser behavior was carried out by M.Kh. M.Ka. B.S., A.M., and Y. F. The manuscript was written by M.Kh with contributions from A.M., M.Ka., B.S., A.S., and Y.F.
16